\documentclass[11pt,twocolumn]{aastex63}
\usepackage{url}
\usepackage{gensymb}
\usepackage{textcomp}
\usepackage{xspace}
\usepackage{natbib,amsmath} \usepackage{graphicx}
\usepackage{hyperref} \usepackage{float} \usepackage{subfigure}
\usepackage{tabularx}
\providecommand{\e}[1]{\ensuremath{\times 10^{#1}}}
\usepackage{bm}

\newcommand{\ang}{\AA\xspace}

\newcommand{\g}{\mathrm{g}}
\newcommand{\K}{\mathrm{K}}

\newcommand{\s}{\mathrm{s}}

\newcommand{\p}{\mathrm{p}}

\usepackage{array}

\begin{document}
\title{Detection of Atmospheric Escape from Four Young Mini Neptunes}

\correspondingauthor{Michael Zhang}
\email{mzzhang2014@gmail.com}

\author[0000-0002-0659-1783]{Michael Zhang}
\affiliation{Department of Astronomy, California Institute of Technology, Pasadena, CA 91125, USA}

\author[0000-0002-5375-4725]{Heather A. Knutson}
\affiliation{Division of Geological and Planetary Sciences, California Institute of Technology}

\author[0000-0002-8958-0683]{Fei Dai}
\affiliation{Division of Geological and Planetary Sciences, California Institute of Technology}

\author[0000-0002-6540-7042]{Lile Wang}
\affiliation{Kavli Institute for Astronomy and Astrophysics, Peking University, Beijing 100871, China}

\author[0000-0003-2058-6662]{George R. Ricker}
\affiliation{Kavli Institute for Astrophysics and Space Research, Massachusetts Institute of Technology}

\author[0000-0001-8227-1020]{Richard P. Schwarz}
\affiliation{Center for Astrophysics, Harvard \& Smithsonian, 60 Garden Street, Cambridge, MA 02138, USA}

\author[0000-0002-9312-0073]{Christopher Mann}
\affiliation{Institute for Research on Exoplanets, University of Montreal}

\author[0000-0001-6588-9574]{Karen Collins}
\affiliation{Center for Astrophysics, Harvard University}

\begin{abstract}
We use Keck/NIRSPEC to survey a sample of of young ($<$1 Gyr), short period mini Neptunes orbiting nearby K dwargs to measure their mass loss via the metastable helium line.  We detect helium absorption from all four of the targets in our initial sample.  The first detection, around TOI 560b, was announced in a previous paper.  We now announce three additional detections around TOI 1430.01, 2076b, and TOI 1683.01.  All four planets show an average in-transit excess absorption of 0.7--1.0\%.  However, the outflows differ in their kinematic properties.  TOI 1430b exhibits pre-ingress absorption, while TOI 2076b's outflow is exceptionally optically thick and shows significant post-egress absorption.  For all four planets, the width of the measured helium absorption signal is consistent with expectations for a photoevaporative outflow (10--30 km/s, 5000--10,000 K).  Unless broadening mechanisms other than thermal velocity and the bulk outflow velocity are significant, our observations disfavor core-powered mass loss models, which predict much slower (1-3 km/s) outflows.  We utilize both an isothermal Parker wind model and an order-of-magnitude method to estimate the mass loss timescale, and obtain $\sim$ a few hundred Myr for each planet.  We conclude that many, if not all, of these planets will lose their hydrogen-rich envelopes and become super Earths.  Our results demonstrate that most mini Neptunes orbiting sun-like stars have primordial atmospheres, and that photoevaporation is an efficient mechanism for stripping these atmospheres and transforming these planets into super Earths.
\end{abstract}

% #####################################################################

\section{Introduction}
\label{sec:introduction}
Planets between 1--4 $R_\Earth$ are frequently found on close-in orbits around nearby sun-like stars \citep{zhu_2021}.  This class of planets has no comparable analogue in the solar system, and there is considerable debate about their origins \citep{drazkowska_2022}.  A now-famous ``radius gap'' or ``Fulton gap'' \citep{fulton_2017,fulton_2018} separates the smaller, denser super Earths (1--1.7 R$_\Earth$) from the larger, less dense mini Neptunes (2--3 R$_\Earth$).  The conventional explanation for this gap is that mini Neptunes have a primordial hydrogen/helium atmosphere comprising $\sim$1\% of their total mass that significantly inflates their radii, whereas super Earths have either lost their hydrogen-rich atmospheres or never acquired them in the first place (e.g., \citealt{lee_2021,lee_2022})).  If most super-Earths initially formed with hydrogen-rich envelopes, they could have been stripped away by intense X-ray and extreme UV (XUV) irradiation from the young star (photoevaporative mass loss, e.g. \citealt{owen_2017,mills_2017}).  Alternatively, the energy powering the outflow may predominately come from the internal heat of formation (`core-powered' mass loss; \citealt{ginzburg_2018,gupta_2019}).  A third possibility is that some or most mini Neptunes have no hydrogen-rich envelopes at all, but instead formed with substantial water-rich envelopes (e.g. \citealt{mousis_2020}).  In order to differentiate between these competing hypotheses, it is crucial to obtain observations of young mini Neptunes to ascertain whether and how quickly they are losing hydrogen/helium.

Even though the study of escaping mini-Neptune atmospheres is in its infancy, observations of escaping atmospheres around larger planets have a long history, as do attempts to understand the observations with models.  The first detection of Ly$\alpha$ absorption occurred two years after the detection of the first transiting exoplanet \citep{vidal-madjar_2003}.  Because the Ly$\alpha$ core is wiped out by the interstellar medium and only the high-velocity wings can be seen, Ly$\alpha$ observations are only logarithmically sensitive to the mass loss rate \citep{owen_2023}.  Helium absorption observations hold more promise because interstellar absorption is not a problem, and because the line probes the inner regions of the outflow.  The 1083 nm helium triplet is a sensitive probe of escaping atmospheres \cite{spake_2018, oklopcic_2018}, and can be readily accessed using ground-based telescopes.  This technique works best for planets with K dwarf hosts, as these stars are predicted to produce the largest metastable helium population \citep{oklopcic_2019}.  Escaping helium has been detected around numerous giant planets, including WASP-107b \citep{spake_2018,allart_2019}, WASP-69b \citep{nortmann_2018}, and HD 189733b \citep{salz_2018,guilluy_2020,zhang_2022}.  Models capable of predicting helium absorption include the Parker wind models of \cite{oklopcic_2018} and \cite{lampon_2020}), multiple 1D hydrodynamic models (e.g. The PLUTO-CLOUDY Interface, or TPCI; \citealt{salz_2015}), the 3D Monte Carlo simulation EVaporating Exoplanets \citep{bourrier_2013b,allart_2018,allart_2019}, and many grid-based 3D hydrodynamic codes \citep{wang_2018,shaikhislamov_2021,khodachenko_2021,macleod_2022,rumenskikh_2022}.  Although these models have been used to reproduce the observed outflows from giant exoplanets, they are largely untested in the mini-Neptune regime.

Earlier this year, we reported the first detections of ongoing mass loss from young mini Neptunes.  In \cite{zhang_2021b}, we detected Ly$\alpha$ absorption from the 400 Myr mini Neptune HD 63433c, which orbits a sun-like star, but not from the inner mini Neptune in the same system.  This shows that the outer planet has an escaping hydrogen/helium atmosphere, but further interpretation is difficult because the interstellar medium blocks the core of the line, and because the non-detection from the inner planet could be due to the absence of a lightweight atmosphere, a magnetic field preventing the outflow from being accelerated to high velocities, or photoionization of escaping hydrogen \citep{owen_2023}.  In \cite{zhang_2021c}, we reported metastable helium absorption from the outflowing atmosphere of TOI 560b, a 500 Myr mini Neptune orbiting a K dwarf.  

In these studies, our Microthena and TPCI models both struggled to reproduce the observations.  The Microthena models could roughly match the depth of the Ly$\alpha$ blue wing absorption for HD 63433c, but could not explain either the red wing absorption or the non-detection around HD 63433b.  For TOI 560b, both the fiducial Microthena and TPCI models overestimated the observed helium absorption, and the Microthena models predicted a blueshift instead of the observed redshift.  We suspect that the poor fit could be due to large uncertainties in planetary and stellar parameters, such as the EUV flux, the atmospheric metallicity, and the stellar wind conditions.  It could also be due to limitations of the model, such as the neglect of magnetic fields, which could significantly affect the outflow (e.g. \citealt{owen_2014,khodachenko_2015,arakcheev_2017}).

To gain a solid understanding of if and how mini Neptunes turn into super Earths, a large sample of mini Neptunes with outflow detections is required, spanning a range of radii, masses, ages, and periods.  It is much easier to accumulate a large sample with helium observations than with Ly$\alpha$ because the latter requires \emph{Hubble Space Telescope} observations of very nearby stars that have favorable radial velocities, and we have not found any additional candidate as favorable as HD 63433.  Having a larger sample of helium detections will reveal whether the features we detected in TOI 560b--such as the existence of an escaping hydrogen/helium atmosphere, the redshift of the signal, the asymmetry in the light curve, or the unexpectedly low amplitude of the absorption--are common or rare.  Population studies can also reveal patterns in the data that give insight into the physical processes involved.  By testing our models on a population level, we can mitigate some of the large uncertainties in individual planet and stellar parameters.  By focusing on young planets which nevertheless span a factor of a few in age, we catch young mini Neptunes during the most critical period of their lives--the period when they could be rapidly losing mass and transitioning into super Earths.

For these reasons, we undertook a survey of escaping helium observations from eight young ($<1$ Gyr) mini Neptunes orbiting K-type stars observable from the Keck telescopes.  This paper presents the first four results from the survey.  All of our targets are recent discoveries by the Transiting Exoplanet Survey Satellite (TESS).  TESS has surveyed the entire sky, finding many more nearby transiting planets than previously known and making a survey of this type possible for the first time.  Section \ref{sec:observations} presents the observations and data reduction, section \ref{sec:results} presents the results, section \ref{sec:mass_loss_rate} estimates the mass loss rate using two different methods and comments on the implications, and section \ref{sec:conclusion} concludes.

\section{Observations and Data Reduction}
\label{sec:observations}
\subsection{Targets}
\begin{table*}[ht]
  \centering
  \caption{Planet and star properties}
  \begin{tabular}{C C C C C}
  \hline
  	  \text{Parameter/Planet} & \text{TOI 560b}^a & \text{TOI 1430.01} & \text{TOI 1683.01} & \text{TOI 2076b}^b\\
      \hline
      R_p (R_\Earth) & 2.79 \pm 0.1 & 2.1 \pm 0.2^d & 2.3 \pm 0.3^d &  2.52 \pm 0.04\\
      M_p (M_\Earth) & 10.2_{-3.1}^{+3.4} & 7 \pm 2?$^f$ & 8 \pm 2?$^f$ &  9?$^f$\\
      T_{\rm eff, *} & 4511 \pm 110 & 5067 \pm 60^c & 4539 \pm 100^c & 5200 \pm 70\\
      R_* (R_\Sun) & 0.65 \pm 0.02 & 0.784_{-0.014}^{+0.018} & 0.636_{-0.017}^{+0.031} & 0.762 \pm 0.016\\
      M_* (M_\Sun) & 0.73 \pm 0.02 & 0.85 \pm^j 0.10 & 0.69 \pm 0.09^j & 0.824_{-0.037}^{+0.035} \\
      P_{\rm rot,*} & 12.08 \pm 0.11 & 5.79 \pm 0.15 & 11.3 \pm 1.5 & 6.84 \pm 0.58\\
      J_* & 7.6 & 7.6 & 8.8 & 7.6\\
      \text{Age (Myr)} & 480-750 & 165 \pm 30^g & 500 \pm 150^g & 204 \pm 50\\
      a (AU) & 0.0604 & 0.0705 & 0.036 & 0.0631\\
      F_{XUV} (erg s$^{-1}$ cm$^{-2}$) & 5000 & 6800 & 12,000 & 9500\\
      F_{MUV} (erg s$^{-1}$ cm$^{-2}$) & 3000 & 27,000 & 4,900 & 50,000\\
      T_{eq} (K) & $714 \pm 21$ & 813 & 927 & 797 \pm 12\\
      P (d) & 6.3980420_{-0.0000062}^{+0.0000067} & 7.434162^d \pm 1.5\e{-5} & 3.05752578^d \pm 6.3\e{-6} & 10.355183 \pm 0.000065^i\\
      T_0 (BJD_{\rm TDB} - 2,457,000) & 1517.69013_{-0.00059}^{+0.00056} & 2032.74865 \pm 0.0010 & 2232.2348 \pm 0.0011 & 2654.9837 \pm 0.0015^i\\
      a/R_* & 19.98_{-0.63}^{+0.61} & 17.25 \pm 0.54 & 9.99 \pm 0.39 & 17.8 \pm 0.4\\
      i($^\circ$) & 88.37 \pm 0.18 & 88.69_{-0.23}^{+0.30} & 85.99_{-0.28}^{+0.29} & 88.9 \pm 0.11$^h$\\
      \text{Transit duration } \tau_{14} (h) & 2.143_{-0.027}^{+0.029} & 2.71_{-0.04}^{+0.07} & 1.28_{-0.04}^{+0.06} & 3.251 \pm 0.03\\
      \hline
  \end{tabular}
  \\
  Sources: $^a$\cite{barragan_2022}, $^b$\cite{osborn_2022}, $^c$ \cite{gaia_2018}, $^d$Own fits to TESS data, $^f$Mass-radius relation of \cite{wolfgang_2016}, $^g$ Gyrochronology using TESS-derived rotation period, $^h$ \cite{hedges_2021}, $^i$ TOI 2076 has TTVs.  See Appendix \ref{sec:appendix_ttvs}, $^j$ TESS Input Catalog v8.2
  \label{table:planet_properties}
\end{table*}

The \emph{Transiting Exoplanet Survey Satellite} (\emph{TESS}) has already surveyed nearly the entire sky, discovering many nearby transiting planets amenable to atmospheric characterization.  The TESS Object of Interest (TOI) catalog \citep{guerrero_2021} contains a mix of new planet candidates and confirmed planets, including previously known planets re-detected by TESS.  To select targets for our Keck survey, we downloaded the catalog from ExoFOP (DOI: 10.26134/ExoFOP5) and imposed several cuts.  We required planet candidates to have radii between 2 and 3 $R_\Earth$, periods less than 20 days,  declinations greater than $-20^\circ$, and K dwarf host stars ($3900 < T_{\rm eff} < 5300$ K) with $J < 9$.  For unpublished candidates, we required a rating of `Validated Planet Candidate?' (VPC?) or higher by the TESS Follow-up Observing Program (TFOP) in order to eliminate likely false positives from our sample.  This designation means that follow-up observations have at least tentatively confirmed that the transit occurs for the correct star, and that there are no Gaia DR2 stars contaminating the TESS aperture that are bright enough to cause the detection.

Most of the planets and planet candidates we identified did not have well-determined ages, so we ran a Lomb-Scargle peridogram on the TESS light curve and required a robustly detected rotation period in order to limit our sample to young stars.  In practice, this means that the star's rotation period must be less than 15 days.   We manually examined the TESS light curve to make sure that the rotational modulations looked strong and convincing (coherent and on the order of tens of percent).  We also searched for the stars in the ROSAT All-Sky Survey, an X-ray catalog.  Detectable X-ray emission is also evidence of youth, and we therefore prioritized targets with detections in this catalog for our survey.  In total, we are currently targeting 8 planets orbiting 7 stars, of which 3 have ROSAT detections.  Although we continue to monitor the list of newly announced TESS candidates, we do not expect the survey sample to expand significantly because TESS has already completed its baseline survey and has likely discovered all of the most favorable targets.  There are additional favorable candidates remaining in the southern hemisphere, but these are not accessible from Keck.

During the first year of the survey, we observed four targets: TOI 560b, 1430.01, 1683.01, and 2076b.  Their properties are summarized in Table \ref{table:planet_properties}.  TOI 560b and 2076b are both confirmed planets with published discovery papers (\citealt{barragan_2022,el_mufti_2021} and \citealt{hedges_2021}, respectively).  They are both the inner planets of multi-planetary systems, with one and two outer mini-Neptune companions respectively.  TOI 1430 and 1683 are unpublished candidates, with the former included in the ROSAT All-Sky Survey, indicating its likely youth.  Neither have any detected transiting companions.  Of the four targets, only TOI 560b currently has a published radial velocity mass measurement.  The masses of the other planets are derived from the mass-radius relation of \cite{wolfgang_2016}.  It is possible that this relation overestimates masses for young planets.  TOI 560b gives us some reason for optimism, because the RV-derived mass of 10.2$_{-3.4}^{+3.4}$ M$_\Earth$ is perfectly consistent with the mass inferred from the mass-radius relation, $10 \pm 2$ M$_\Earth$.  Similarly, the relation gives $8.4 \pm 2$ for TOI 560c, compared to an RV mass of $9.7_{-1.7}^{+1.8}$ M$_E$.  We encourage RV followup of the other planets to aid the interpretation of atmospheric observations.

For TOI 1430.01 and 1683.01, which do not have published discovery papers, we took additional steps to validate their planetary nature.  We checked that the even transits had a consistent transit depth with the odd transits.  TOI 1683.01 was verified (in part) by the Dragonfly Telephoto Array, a robotic telescope stationed in New Mexico that is comprised of 48 individual telephoto lenses acting in concert similar to a 1.0-m refractor.  With Dragonfly, we detected an on-time and on-target 1.6 ppt transit event, excluding the nearby Gaia stars as potential causes of a false-positive detection by TESS.  Photometric collected by the Las Cumbres Observatory have only tentatively detected the transit of TOI 1430.01, due to its shallow transit depth.  We searched the Gaia DR2 catalog to ensure that no bright stars are within 21\arcsec\xspace{} (one TESS pixel) of the target, as this could contaminate the light curve.  For TOI 1430.01, Gaia detected 6 stars within this radius, but their total flux in the Gaia G band is only 0.03\% that of TOI 1430.  Near TOI 1683, Gaia detected one star with a flux 0.025\% that of TOI 1683.  Gaia has an angular resolution of 0.4\arcsec, corresponding to 17 AU at the distance of TOI 1430 and 21 AU at the distance of TOI 1683.  We therefore conclude that, when combined with our detection of helium absorption, these tests suggest that these two candidates are likely to be real planets and treat them as such in the subsequent analysis.

For these two planets, we obtained transit parameters by fitting the TESS light curves using the same methodology as \cite{dai_2021}.  To summarize, we downloaded the TESS photometry from the Mikulski Archive for Space Telescopes (MAST).  We fit each light curve with \texttt{batman} \citep{kreidberg_2015}, with free parameters including the out-of-transit flux level, stellar density, quadratic limb darkening coefficients, period, transit midpoint, $R_p/R_s$, and orbital inclination.  For both planets, we have multiple sectors of TESS observations spanning 2 years, giving us a very accurate ephemeris that is sufficient to predict transit times corresponding to our helium observations to within 2--3 minutes.  The phased light curves and transit fits are plotted in Appendix \ref{sec:tess_light_curves}.

We obtained age estimates for TOI 1430 and 1683 using gyrochronology.  Specifically, we used \cite{schlaufman_2010}, which relates age to stellar mass and rotation period, and \cite{mamajek_2008}, which relates age to B-V color and rotation period.  We obtained the rotation period from a Lomb-Scargle periodogram of the TESS SAP fluxes, and take the masses and B-V colors from the TESS Input Catalog \citep{stassun_2019}.  These stellar masses are computed from the effective temperature, which in turn is obtained from spectroscopy (if available) or deredenned colors.  For TOI 1430, we obtain $156 \pm 30$ Myr with the first method and $174 \pm 8$ Myr using the second; for TOI 1683, we obtain $550 \pm 150$ and $460 \pm 110$ Myr.  We adopt the average of the two methods and the more conservative error bars.

\subsection{Keck/NIRSPEC}
\label{subsec:keck_reduction}
We used Keck/NIRSPEC to observe one transit for each of the four planets.  All spectra were observed in $Y$ band with the 12 $\times$ 0.432$\arcsec$ slit.  The exposure time was 60 seconds per frame, and we used an ABBA nod to subtract background, giving us a typical observing efficiency of 77\%.

\begin{table*}[ht]
  \centering
  \caption{Keck/NIRSPEC observations}
  \begin{tabular}{c c c c c}
  \hline
  	  Parameter/Planet & TOI 560b & TOI 1430.01 & TOI 1683.01 & TOI 2076b\\
      \hline
      Date (UTC) & 2021-03-18  & 2021-09-21  & 2021-10-01  & 2022-03-16 \\
      Time (UTC) & 12:06--16:07 & 05:04--09:58 & 10:31--15:51 & 12:06--16:07\\
      SNR & 150 & 160 & 67 & 190\\
      Pre-ingress baseline (h) & 0.7 & 0.7 & 2.4 & 0\\
      Transit fraction observed (\%) & 100 & 100 & 100 & 31  \\
      In-transit duration (h) & 2.1 & 2.7 & 1.3 & 1.0 \\
      Post-egress baseline (h) & 0.6 & 1.4 & 1.6 & 3.0\\
      \hline
  \end{tabular}
  \label{table:nirspec_data}
\end{table*}

Table \ref{table:nirspec_data} shows the target-specific details for each observation.  We include TOI 560b for comparison, with the full details for this target provided in \cite{zhang_2021c}.  All four nights of observations were affected to varying degrees by weather.  During the observations of TOI 560b, 1430.01, and 1683.01, transparency was poor throughout the night.  The 1683.01 half-night also suffered from poor seeing.  When we began observing at the beginning of the night the seeing was 1\arcsec, but it started rising around 12:44 UTC, eventually reaching values as high as 3\arcsec.  The combination of a faint host star, clouds, and slit losses from poor seeing resulted in a low SNR for this target.  For TOI 2076b, a combination of telescope problems, fog, and unexpected transit timing variations (see Appendix \ref{sec:appendix_ttvs}) prevented us from observing the first 2/3 of the transit.  However, the remainder of the night had stable seeing and no detectable clouds.

To analyze the data, we used the same pipeline described in \cite{zhang_2021c}.  Among other steps, this pipeline produces A-B subtraction images to remove the background, uses optimal extraction on the differenced images to obtain 1D spectra, fits a template to the spectra to obtain wavelength solutions, and runs \texttt{molecfit} to remove telluric absorption features.  The pipeline calculates excess absorption during the transit as a function of time and wavelength, which can then be collapsed along either axis to give the average in-transit excess absorption spectrum or the band-integrated light curve.

In addition to fitting telluric features, \texttt{molecfit} also fits for the full width half maximum (FWHM) of the line spread profile, but it is not stable enough to reliably fit the FWHM for every spectrum.  Instead, we run it on every spectrum and take the median of the FWHMs across the night for every target except TOI 1683, which has low SNR.  We obtain 3.4 pixels, corresponding to R=32,000 at the position of the helium line.  We then rerun all the \texttt{molecfit} fits with the FWHM fixed to 3.4 pixels.  This value is close to what we obtain for each individual night: $3.65 \pm 0.38$ (TOI 560), $3.36 \pm 0.17$ (TOI 1430), $3.40 \pm 0.12$ (TOI 2076), where the reported error is the standard deviation of the FWHM across the night.

\begin{figure*}
  \centering 
  \subfigure {\includegraphics[width=0.5\textwidth]{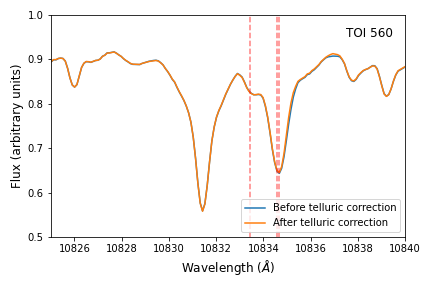}}\subfigure {\includegraphics[width=0.5\textwidth]{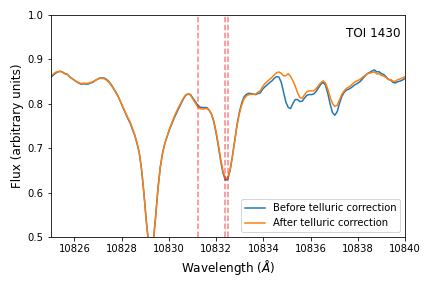}}
  \subfigure {\includegraphics[width=0.5\textwidth]{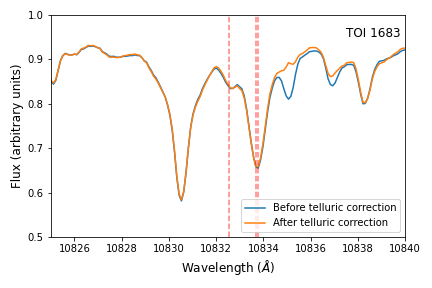}}\subfigure {\includegraphics[width=0.5\textwidth]{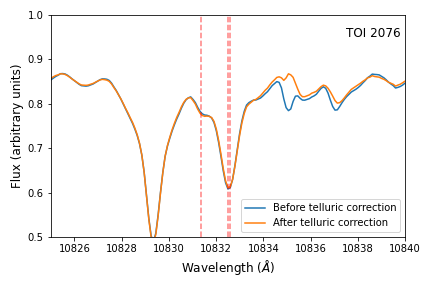}}
  \caption{The stellar spectra in the vicinity of the helium triplet (in red), before and after correction for tellurics.  Unlike in the rest of the paper, wavelengths are in Earth's frame, not the stellar rest frame, leading telluric lines to have the same wavelengths for every target.  The telluric lines do not overlap with the helium line for any target except TOI 560, for which we were lucky to have had an exceptionally dry night.}
\label{fig:tellurics}
\end{figure*}

Figure \ref{fig:tellurics} shows the stellar spectrum in the vicinity of the helium line, both before and after correction for tellurics by \texttt{molecfit}.  The strongest stellar line is a Si I at a rest frame vacuum wavelength of 10830\ang; the second strongest line is the metastable helium line, at 10830\ang.  For TOI 1430 and 2076, the helium line was far from any tellurics.  For TOI 1683, telluric absorption begins to pick up only at the reddest edge of the helium line, 20 km/s from the center.  As we will see, the observed planetary absorption signal is blueshifted, making the telluric absorption even less important.  It is only for TOI 560 that a telluric line falls right on top of the helium line.  Due to extraordinary luck, telluric absorption was minimal because this night happened to be exceptionally dry--\texttt{molecfit} reports a precipitable water vapor (PWV) column of 0.15 mm, compared to 0.9 mm for TOI 1430, 1.0--1.4 mm for TOI 2076, and $\sim$1 mm for TOI 1683.  We discuss the telluric correction for TOI 560 in greater detail in \cite{zhang_2021c}, and are confident that it does not greatly affect our conclusions.

\subsection{X-ray observations}
We also separately observed all of the host stars except for TOI 2076 using XMM-Newton to characterize their X-ray spectra, which are needed as inputs for photoevaporative mass loss models.  XMM-Newton observes with all six instruments simultaneously, of which the most useful for our purposes are the three EPIC cameras and the Optical Monitor (OM).  The OM observed with the UVW2 and UVM2 filters in order to measure the star's mid-ultraviolet flux, which destroys metastable helium.  We followed the procedure outlined in \cite{zhang_2021b} to compute the star's X-ray spectrum and MUV flux from the XMM-Newton EPIC and OM observations, respectively, except that we only fit a one-temperature model for TOI 1683 due to its much lower S/N.  The observation IDs, along with the X-ray spectra, light curves, and inferred model parameters for TOI 1430 and 1683 are given in Appendix \ref{sec:appendix_xrays}.  The corresponding X-ray observations for TOI 560 are described in \cite{zhang_2021c}.

TOI 2076 does not have a XMM-Newton observation from our program, but it will be observed as part of PI Wheatley's program 090431 in the current cycle.  TOI 2076 was also observed by the ROSAT All-Sky Survey \citep{boller_2016}.  However, RASS only measured 28 photons, too few to fit the spectrum with a plasma model and obtain meaningful constraints.  We instead assume that TOI 2076 has the same X-ray spectral shape as TOI 1430, a star whose estimated age and effective temperature are within $1\sigma$ of that of TOI 2076.  We scale TOI 1430's X-ray flux to 17 erg s$^{-1}$ cm$^{-2}$ at 1 AU, derived from the $L_X/L_{\rm Bol}$ for TOI 2076 obtained from ROSAT data by \cite{hedges_2021}.

\section{Results}
\label{sec:results}
\begin{figure*}
  \centering 
  \subfigure {\includegraphics[width=0.5\textwidth]{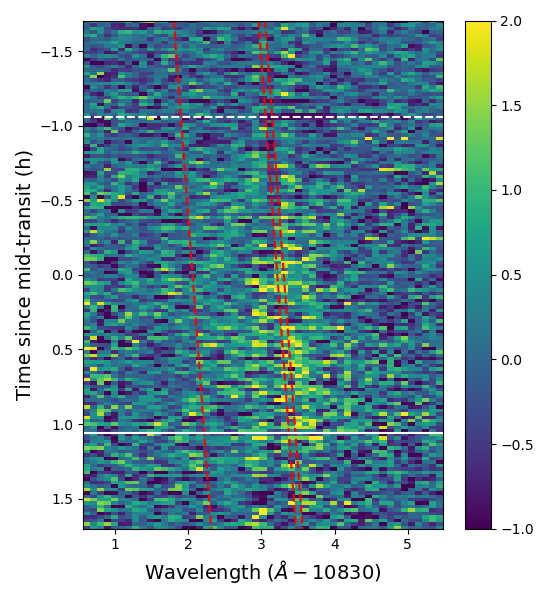}}\subfigure {\includegraphics[width=0.5\textwidth]{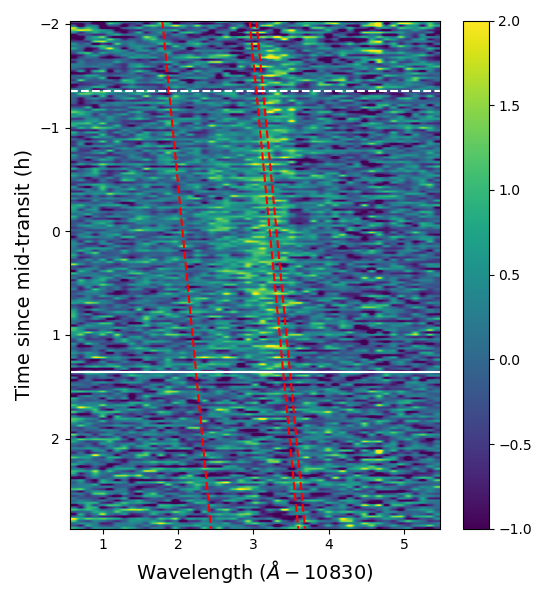}}
  \subfigure {\includegraphics[width=0.5\textwidth]{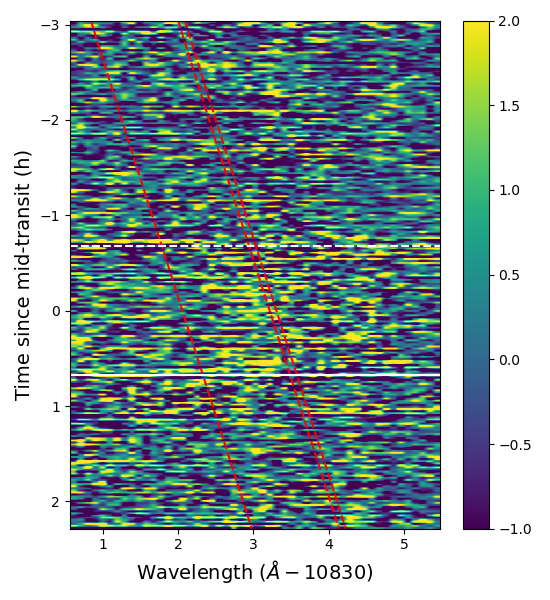}}\subfigure {\includegraphics[width=0.5\textwidth]{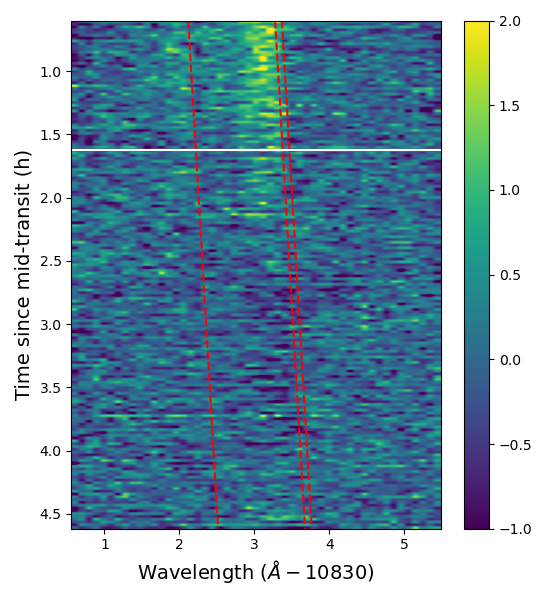}}
  \caption{Excess absorption, in percent, as a function of time and wavelength in the stellar rest frame.  Clockwise from upper left: TOI 560b (first published in \citealt{zhang_2021c}), 1430.01, 2076b, 1683.01.  The colorbars and wavelengths are matched, but the y axes are not.  The red lines show the wavelengths of the helium triplet in the planet rest frame.  The horizontal dashed and solid lines show the beginning of white light ingress and end of white light egress, respectively.}
\label{fig:heatmaps}
\end{figure*}

\begin{figure*}
  \centering 
  \subfigure {\includegraphics[width=0.5\textwidth]{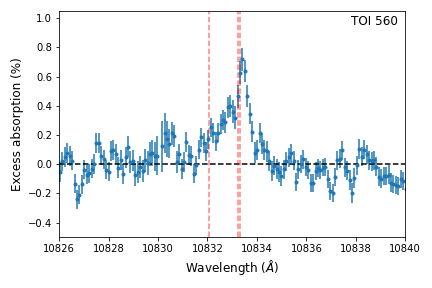}}\subfigure {\includegraphics[width=0.5\textwidth]{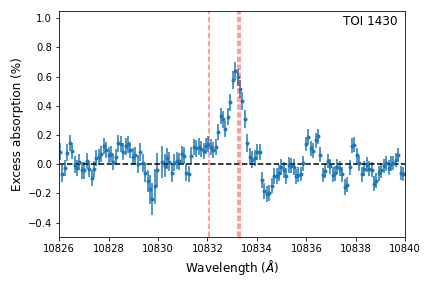}}
  \subfigure {\includegraphics[width=0.5\textwidth]{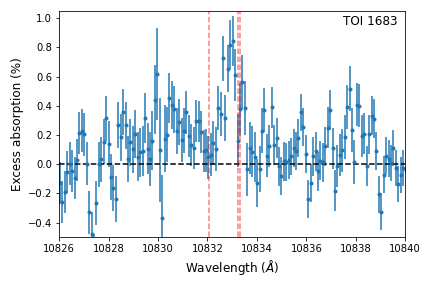}}\subfigure {\includegraphics[width=0.5\textwidth]{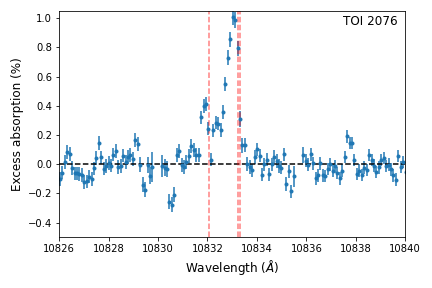}}
  \caption{Average in-transit excess absorption spectrum.  Note that we only have the last third of the transit for TOI 2076, making it difficult to directly compare to the other full-transit observations.}
\label{fig:excess_1D}
\end{figure*}

\begin{figure*}
  \centering 
  \subfigure {\includegraphics[width=0.5\textwidth]{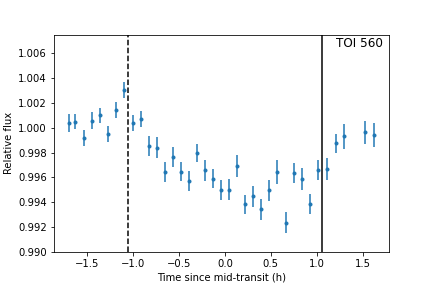}}\subfigure {\includegraphics[width=0.5\textwidth]{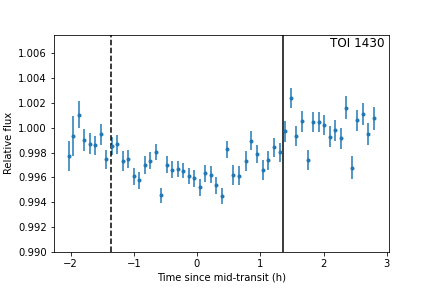}}
  \subfigure {\includegraphics[width=0.5\textwidth]{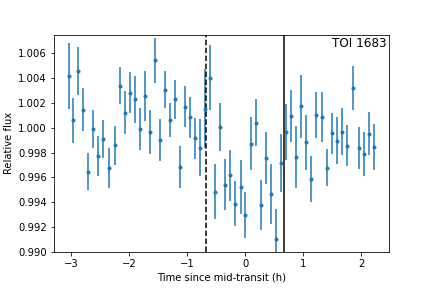}}\subfigure {\includegraphics[width=0.5\textwidth]{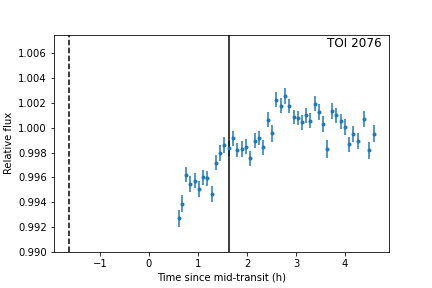}}
  \caption{Band-integrated light curves, integrated within 1.5\ang of the main peak of the helium triplet, at 10833.3\ang.  The wavelength range is defined with respect to the stellar frame in order to give an accurate picture of stellar variability in the helium line.  The y axes are matched, but not the x axes.  Dashed and solid vertical lines represent the beginning and end of the white light transit, respectively.}
\label{fig:light_curves}
\end{figure*}

% \begin{table*}[ht]
%   \centering
%       \caption{Excess absorption spectra properties}
%   \begin{tabular}{c C C C C}
%   \hline
%   	  Planet & \text{Redshift (km/s)} & \text{Ratio} & \text{Height (\%)} & \text{Width (\ang)} \\
%       \hline
%       TOI 560b    & 2.2_{-1.8}^{+1.6}  & 0.33 \pm 0.10 & 0.54 \pm 0.06 & 0.44 \pm 0.06\\
%       TOI 1430.01 & -5.4_{-1.2}^{+1.0} & 0.22 \pm 0.07 & 0.56 \pm 0.05 & 0.37 \pm 0.03\\
%       TOI 1683.01 & -9.4_{-2.7}^{+3.1} & 0.31_{-0.17}^{+0.20} & 0.62 \pm 0.14 & 0.40_{-0.08}^{+0.10}\\
%       TOI 2076b   & -7.0 \pm 0.07 & 0.33_{-0.06}^{+0.10} & 0.97 \pm 0.08 & 0.26 \pm 0.02\\
%       \hline
%   \end{tabular}
%   \label{table:gaussian_fits}
% \end{table*}

\begin{table*}[ht]
  \centering
      \caption{Excess absorption spectra properties}
  \begin{tabular}{c C C C C C}
  \hline
  	  Planet & \text{Height (\%)} & \text{Redshift (km/s)} & \text{Ratio*} & \text{Equivalent Width (m\ang)} & \text{EW/Height (\ang)} \\
      \hline
      TOI 560b    & 0.72 \pm 0.08 &  4.3 \pm 1.4 & <0.27 \pm 0.20 & 8.6 \pm 0.6 & 1.19 \pm 0.16\\
      TOI 1430.01 & 0.64 \pm 0.06 & -4.0 \pm 1.4 & <0.19 \pm 0.09 & 6.6 \pm 0.5 & 1.03 \pm 0.12 \\
      TOI 1683.01 & 0.84 \pm 0.17 & -6.7 \pm 2.8 & <0.32 \pm 0.11 & 8.5 \pm 1.6 & 1.01 \pm 0.28\\
      TOI 2076b   & 1.01 \pm 0.05 & -6.7 \pm 1.4 & <0.40 \pm 0.05, >0.18 \pm 0.05 & 10.0 \pm 0.7 & 0.99 \pm 0.09\\
      \hline
  \end{tabular}
  \\
  $^*$ Only TOI 2076b has a visually distinguishable secondary peak.\\
  \label{table:spectra_properties}
\end{table*}
%dm/dt = eta*pi R^3 F_XUV / (GM)
Figure \ref{fig:heatmaps} shows the excess absorption from each of the four planets as a function of time and wavelength, relative to the average flux outside of the white light transit.  Figure \ref{fig:excess_1D} shows the average in-transit excess absorption spectrum in the planetary rest frame, while Figure \ref{fig:light_curves} shows the band-integrated light curve in a 3\ang bandpass centered on the main helium peak.

The four absorption signals share both striking similarities and striking differences.  Despite having different radii, orbital periods, ages, and masses, all four planets show an average in-transit absorption depth of 0.7--1.0\%.   None of the planets show an evolving velocity that tracks the radial acceleration of the planet.  This might cause one to suspect that stellar activity is the origin of the putative planetary signals, but the band-integrated light curves strongly argue for a planetary origin: they decrease at the white light ingress and recover by the white light egress for all planets except TOI 2076b.  One final common feature of all our observations is unexplained variations in the light curve, such as the rise prior to ingress for TOI 560b, the dip just after ingress for TOI 1430.01, the dip well before ingress for TOI 1683.01, and the decline for TOI 2076b 3 hours after egress.  We attribute these features to stellar activity, because our targets are young stars and the stellar helium line is a well known tracer of chromospheric activity.  It is also conceivable that some of this variability is due to absorption from the escaping planetary atmosphere, which may form extended and complex structures (see e.g. the 3D simulations of \citealt{mccann_2019}).  

The differences between the planets are as interesting as the similarities.  One of the most puzzling features of the TOI 560b absorption signal, and the hardest to explain with our 3D models, was the redshift \citep{zhang_2021c}.  Young stars are expected to have a strong stellar wind pushing the outflow away from the star, which is toward the observer during transit, and previous helium observations of giant exoplanets--including HD 189733b \citep{salz_2018,guilluy_2020,zhang_2022}, WASP-107b \citep{allart_2019,kirk_2020}, WASP-69b \citep{nortmann_2018}, GJ 3470b \citep{palle_2020}, HD 209458b \citep{alonso-floriano_2019}, and HAT-P-11b \citep{allart_2019}--saw a blueshift.  The sole exception is HAT-P-32b \citep{czesla_2022}.   Our new observations show that the redshift we observed was unusual, as our other three planets all show net blueshifts.  Also potentially exceptional with TOI 560b was the time evolution of the absorption; in both Figure \ref{fig:heatmaps} and \ref{fig:light_curves}, it is evident that absorption peaks after mid-transit.  TOI 1430.01 does not show this characteristic, but more data is required to evaluate whether TOI 1683.01 and 2076b do.

If TOI 560b has its quirks, the other planets are no less peculiar.  TOI 1430.01 shows what appears to be pre-ingress absorption.  Without any detected velocity evolution matching the planets orbit, it is difficult to determine whether this pre-ingress absorption might be due to stellar variability, or a combination of planetary absorption and stellar variability.  Real pre-ingress absorption is not implausible, as it has been seen for HAT-P-32b \citep{czesla_2022} and can be produced by an up-orbit stream (e.g. \citealt{lai_2010,mccann_2019,matsakos_2015}).  We are not aware of any published 3D hydrodynamic simulations that predict strong pre-ingress helium absorption, but \cite{macleod_2022} predicts weak pre-ingress absorption in their simulation with an intermediate-strength ($\sim$10x solar) stellar wind.  Incidentally, in their simulation with a strong stellar wind ($\sim$100x solar), the helium absorption has considerably less velocity evolution than the planet itself--reminiscent of our observations.  Finally, TOI 2076b is unique among our four planets in showing extensive absorption for at least half an hour after white light egress, with the band-integrated light curve only declining 50 minutes after egress.  

TOI 2076b is also unique in showing both a clear secondary peak and a primary peak, both blueshifted.  The peak ratio is $0.40 \pm 0.05$, and there is a distinct valley in between the two peaks.  The peak ratio is 0.125 in the case of a perfectly optically thin outflow and 1 for a perfectly optically thick outflow; thus, the observed peak ratio indicates an effective optical depth in between these two extremes.  We do not detect a secondary peak for the other three planets, indicating that their outflows are likely closer to the optically thin limit.  There are relatively few detections of secondary peaks in published helium observations of giant planets (e.g. HD 189733b, HAT-P-32b), and to our knowledge, no other planet has such a prominent secondary peak separated from the main peak by such a prominent valley.  This is partly due to the high SNR of our detection; with higher SNR data, HAT-P-32b \citep{czesla_2022} may turn out to have a similar peak ratio and valley depth.  \cite{salz_2018} report a similar peak ratio of $0.36 \pm 0.03$ for HD 189733b, corresponding to an optical depth in the main peak of 3.2, although the peaks are not as clearly separated.

Coincidentally, \cite{gaidos_2022} observed the same transit of TOI 2076b from the same mountain, using the high resolution InfraRed Doppler (IRD) spectrograph on the 8.2 m Subaru telescope.  They obtain lower S/N than we do, with much higher correlated noise. Their results are consistent with ours: 1\% absorption in the helium line during transit, with some post-egress absorption.  However, they tentatively interpret their results as stellar activity rather than planetary absorption.  Given that we see planetary absorption of similar amplitude and width in every planet of our survey, we disagree with this interpretation, but more transits will be needed to conclusively disambiguate between these two possibilities.

Table \ref{table:spectra_properties} shows the quantitative properties of the outflow: peak absorption, redshift, peak ratio, and equivalent width.  Defining these properties is not trivial because the line profile is unknown.  We attempted to fit two Gaussians to the excess absorption spectrum, one for each absorption peak, but the fit was not good and would have resulted in misleading values if used for quantitative purposes.  Instead, we take a data-driven approach.  The peak absorption is taken from the highest datapoint.  The redshift is similarly taken from the position of the highest datapoint, and assigned an error of one pixel (TOI 1683.01) or half a pixel (all other planets).  The peak ratio is calculated by dividing the mean of the three points closest to the secondary peak by the mean of the three points closest to the primary peak.  The primary peak is identified by the point of highest absorption, while the secondary peak is assumed to be blueshifted or redshifted by the same amount as the primary peak.  Because we do not know how much of the absorption at the secondary peak is actually due to highly blueshifted gas absorbing at its primary peak, these ratios are at all upper limits.  For TOI 2076b, we also obtain a lower limit by subtracting off the depth of the valley. The equivalent width is calculated from the light curves in Figure \ref{fig:light_curves}.  For each planet except TOI 2076b, we reject the two lowest in-transit points to account for outliers, and average the next four lowest in-transit points.  For TOI 2076b, for which we have limited in-transit coverage, we average the two lowest in-transit points.  These are next to each other and are the closest points to mid-transit, giving us increased confidence that the points are not outliers.  Finally, we calculate the width of the absorption by dividing the equivalent absorption by the height.  The width is the most consistent metric, and is within 10\% of 1.1\ang for all planets.

\begin{figure}[ht]
  \includegraphics
    [width=0.48\textwidth]{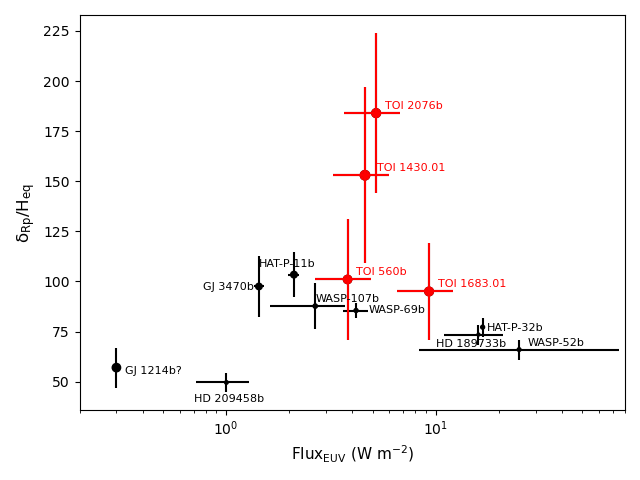}
    \caption{The four mini Neptunes in context, with the size of each point inversely proportional to the planetary radius.  The y-axis shows $\delta_{Rp}/H_{eq}$, the ratio of the increase in apparent radius at 1083 nm to the scale height at equilibrium temperature.  For all of our planets (red filled circles), the y-axis error bars are dominated by the uncertainty in mass.  In particular, planets other that TOI 560 have no mass measurement.  Data for published planets (black filled circles) are taken from \citealt{kasper_2020}, with updates from \cite{orell-miquel_2022} (GJ 1214b, tentative) and \cite{kirk_2022} (WASP-52b).  Photometric detections are excluded because they are not comparable.  Breaking with precedent, upper limits are excluded to reduce clutter.}
\label{fig:scale_heights}
\end{figure}

Finally, Figure \ref{fig:scale_heights} shows the maximum excess absorption in units of scale height and compared to the EUV flux.  This way of comparing planets has been used in many previous papers (e.g. \citealt{nortmann_2018,orell-miquel_2022}).  The uncertainties are large because the planetary masses are not well known, but the mini Neptunes tend to be higher on this metric than the gas giants.

\section{Mass loss rate}
\label{sec:mass_loss_rate}

\begin{table*}[ht]
  \centering
      \caption{Estimated timescale to lose 1\% of total planet mass.  }
  \begin{tabular}{c c C C C C}
  \hline
  	  Planet & OOM (Gyr) & \text{Parker (Gyr)} & T_{\rm Parker} (kK)& \eta_{\rm OOM} & \eta_{\rm Parker}\\
      \hline
      TOI 560b & 0.9 & 0.12_{-0.04}^{+0.05} & 9.8 \pm 0.7 & 0.5 & 3.9\\
      TOI 1430.01 & 0.5 & 0.10 \pm 0.03 & 6.7 \pm 0.3 & 0.5 & 2.8\\
      TOI 1683.01 & 0.7 & 0.61_{-0.27}^{+0.54} & 6.7 \pm 1.4 & 0.2 & 0.3\\
      TOI 2076b & 0.6 & 0.70_{-0.11}^{+0.16} & 5.0 \pm 0.4 & 0.4 & 0.3\\
      \hline
  \end{tabular}\\
      Despite the naming, both methods should be considered order-of-magnitude estimates only. (see Section \ref{sec:mass_loss_rate}).
  \label{table:mass_loss_timescales}
\end{table*}

Our next step is to translate the magnitude of the measured helium absorption signal into a present-day mass loss rate rate for each planet.  Unfortunately, this rate is not directly derivable from the data.  The observed absorption comes from a trace species--helium in the triplet ground state--comprising less than a millionth of the atoms in the outflow, and whose abundance is determined by photoionization and collisional de-excitation, among other processes, and therefore changes with distance from the planet.  The properties of the outflow depend on the metallicity of the atmosphere (e.g. \citealt{zhang_2021c}), which could plausibly range from 1x to 100x solar for mini Neptunes.  The stellar wind and planetary and interplanetary magnetic fields influence the outflow geometry, which in turn alters the magnitude and shape of the absorption signature.  The wind and magnetic field have highly uncertain strengths, especially because the wind is highly variable.  For these reasons and others, we have previously struggled to match our Ly$\alpha$ absorption data for HD 63433 b/c \citep{zhang_2021b} and the helium absorption data for TOI 560b \citep{zhang_2021c} with either the 1D hydrodynamics code The Pluto-CLOUDY Interface \citep{salz_2015} or the 3D hydrodynamics code Microthena \citep{wang_2018}.  Without an accurate physical model of the outflow, it is impossible to derive a precise mass loss rate.

Instead, we calculate order-of-magnitude estimates of the mass loss rate using two methods.  The first method, introduced in \cite{zhang_2021c}, estimates the mass of metastable helium crossing in front of the star from the maximum equivalent width of the absorption, assuming the outflow is optically thin.  It then converts the mass of metastable helium into total mass, assuming that $10^{-6}$ of the helium atoms are in the metastable state, and that 10\% of all atoms are helium while the other 90\% are hydrogen.  The $10^{-6}$ ratio is typical of a K-type host star, as shown by our 1D and 3D simulations of HD 63433 b/c and TOI 560b, and by the independent models of \cite{oklopcic_2019} (their Figure 2).  To obtain a mass loss rate, this total mass is divided by a timescale on which the mass disappears from view, which we take to be the radius of the star divided by a typical sound speed (10 km/s).

The second method we use to estimate the mass loss rate is to put the Parker wind model of \cite{oklopcic_2018} in a Bayesian framework and run nested sampling retrievals on our datasets.  The model requires the stellar spectrum, which we reconstruct using the methodology of \cite{zhang_2021c}.  Importantly, the EUV reconstruction is highly uncertain because the stellar EUV is only measurable from space and no space telescopes currently have EUV capabilities \citep{france_2022}.  For example, \cite{france_2022} compared four different methods of reconstructing the EUV spectrum for Proxima Centauri and found that they were discrepant by 3--100x, depending on the wavelength; an order of magnitude difference in EUV flux can result in a 3x difference in helium absorption depth \citep{oklopcic_2019}.

Leaving aside the EUV uncertainty, the Parker wind forward model has three free parameters: the log of the mass loss rate, the temperature, and the velocity offset, all of which are given uniform priors.  The first two parameters are required to compute the Parker wind, while the third is meant to account for the moderate redshifts and blueshifts seen in the data, which could be due to a combination of the non-spherical nature of photoevaporation (only the dayside is irradiated), the stellar wind, and a small eccentricity of the planetary orbit.  The Parker wind model outputs optical depth as a function of impact parameter and wavelength, which we then combine with the stellar radius, stellar limb darkening coefficients, and transit parameters to derive a prediction of the excess absorption as a function of wavelength and time.  The log likelihood is computed by comparing this prediction to the observed absorption and error, and the nested sampling code \texttt{dynesty} \citep{speagle_2019} repeatedly calls the log likelihood function to compute the posterior distribution.

This model assumes that the outflow is isothermal, that it is spherically symmetric, and that it consists only of hydrogen and helium in a 10:1 ratio.  These assumptions are of dubious validity.  For TOI 560b, our fiducial 3D simulations showed an outflow that rose in temperature from 1000 K at launch to almost 8000 K at the edge of the interface with the stellar wind, while the 1D solar metallicity simulation showed an outflow that declined from 10,000 K at maximum to 3000 K at 15 $R_p$.  The TOI 560b outflow is also not symmetric, as can be seen from Figure \ref{fig:heatmaps} and \ref{fig:light_curves}; and neither is the TOI 1430.01 outflow, if the pre-ingress absorption is not in fact stellar variability.  Lastly, mini Neptunes can plausibly have very high atmospheric metallicities.  For all of these reasons, we do not expect the Parker wind model to be much better at estimating the mass loss rate than the order-of-magnitude method.

All timescale estimates are shown in Table \ref{table:mass_loss_timescales}.  These were obtained by assuming that 1\% of the planetary mass is in the envelope, with the planetary mass being taken from Table \ref{table:planet_properties}.  After estimating the timescale using both methods, we computed the mass loss efficiency, as defined by the $\eta$ in:

\begin{align}
    \frac{dm}{dt} = \frac{\eta}{4} \frac{R_{XUV}^3 L_{XUV}}{GM_p a^2}
    \label{eq:energy_limited}
\end{align}

To roughly calculate $R_{XUV}$, we follow \cite{wang_2018} in assuming that the EUV photosphere is at $\rho=10^{-13}$ g cm\textsuperscript{-3}.  Assuming that the atmosphere is isothermal at the equilibrium temperature, and that the white light radius corresponds to P=100 mbar, we can derive the radius of the EUV photosphere (which we assume to be similar to the XUV photosphere):

\begin{align}
\rho_{\rm phot} &= \frac{P \mu}{k_B T_{eq}}\\
\beta &\equiv \frac{GM_c \mu}{R_p k_B T_{eq}}\\
R_{\rm EUV} &= \frac{R_p}{1 + \beta^{-1} \ln{(\rho_{EUV} / \rho_{\rm phot})}}\\
\end{align}

Adopting this procedure, we find XUV radii 28--43\% higher than the white light radii.  Changing the pressure of the white light photosphere by a factor of 10 only changes this result by a few percent, but violating the isothermal assumption would result in larger changes.

Using Equation \ref{eq:energy_limited}, we calculate the mass loss efficiency for each planet and report it in Table \ref{table:mass_loss_timescales}.  Just like the mass loss timescales, these efficiencies should not be trusted to better than a factor of a few, and efficiencies greater than 1 are particularly unrealistic.  However, taking into account the large uncertainties, our observational values are sensible.  \cite{caldiroli_2022} used the 1D hydrodynamic code ATES to calculate the efficiency of photoevaporation from a variety of planets with different gravitational potentials and XUV irradiations.  For planets similar to ours in gravitational potential and XUV irradiation, they found $\eta \approx 0.7$, comparable to the values we computed from our observations.  Our observations allow us to rule out scenarios in which the planets have a very high gravitational potential, corresponding to very low efficiencies ($\sim3 \times 10^{-4}$).

Our detection of escaping helium from all of the first four planets in our survey strongly suggests that most mini Neptunes have hydrogen/helium atmospheres, disfavoring the water world hypothesis for their low densities.  The high mass loss rates we infer, which are sufficient to strip the envelope in hundreds of Myr, show that the radius gap separating super Earths and mini Neptunes must evolve over a planet's lifetime.  However, they do not necessarily disprove the primordial origin of the radius valley, as \cite{lee_2021} find that photoevaporation would fill in a primordial radius valley instead of deepening it.  Further work is necessary to investigate the consistency of our observations with the primordial radius valley hypothesis.

\subsection{Photoevaporation or core powered mass loss?}
%Add one more caveat about metallicity
It is worthwhile to ask why the data favors high temperatures.  As many authors have pointed out (e.g. \citealt{mansfield_2018,vissapragada_2020}), there is a degeneracy between temperature and mass loss rate when only the equivalent width of the helium absorption is known, but this degeneracy is broken when the line shape is resolved.  Specifically, the width of the line constrains the temperature (e.g. \citealt{dos_santos_2022}).  For a Parker outflow, the sound speed, the bulk outflow speed, and the thermal speed are all close to $\sqrt{\frac{2k_B T}{\mu m_H}}$ = 9.4 km/s for T=7000 K and $\mu=1.3$, leading to a FWHM of $2.355\sigma \approx 22$ km/s.  By contrast, core-powered mass loss, at $T \sim 800$ K, would exhibit a FWHM of only 7.5 km/s.  Absorption from photoevaporation is just wide enough to be resolvable with NIRSPEC, whose line spread profile coincidentally has a FWHM of 9.4 km/s.  With the line width revealing the temperature, the degeneracy is broken, and the model fits provide tight constraints on both the outflow temperature and the corresponding mass loss rate.

Taken at face value, our Parker wind model appears to favor photoevaporation over core-powered mass loss as the cause of the outflows observed for all four planets.  The inferred outflow temperature of several thousand Kelvin is typical of photoevaporative models (see e.g. the simulations of \citealt{salz_2015b}), whereas a core-powered outflow should have a temperature closer to the planet's equilibrium temperature of $\sim$800 K (e.g. \citealt{ginzburg_2018}).  For these planets, it is not possible for a core-powered outflow to set the mass loss rate by going supersonic before EUV photons can penetrate the outflow (e.g. \citealt{bean_2021}) because the sonic radius for an outflow at $T \sim 800$ K is at $\sim 20 R_p$, whereas the helium signal's effective radius is a few $R_p$.  The inferred mass loss timescales from the observed signals are on the order of hundreds of Myr, comparable to the ages of the systems.  This implies that the observed outflows can strip most or all of the envelope within the first Gyr of the planets' lives.

This argument, however, rests on the assumption that the only significant broadening sources are thermal motion and the bulk velocity predicted by the Parker wind model.  While natural broadening is small enough to be negligible--and is in any case included in the model--the same may not be true for the stellar wind, especially in combination with interplanetary and planetary magnetic fields.  Acting together, these mechanisms or others yet unknown are capable of accelerating a small amount of gas to $>$100 km/s, as evidenced by Ly$\alpha$ absorption detected at these velocities (e.g. \citealt{vidal-madjar_2003,des_etangs_2012}.  However, Ly$\alpha$ can typically only probe high velocities because absorption from the interstellar medium wipes out the line core, and it is likely that the observed high-velocity gas comprises only a tiny fraction of the total outflow.  

Whether stellar winds substantially broaden the helium absorption, which probes the outflow properties closer to the planet ($\sim$3 planetary radii vs. $\sim$12 planetary radii), is uncertain.  The blueshift seen for every planet except TOI 560b, the asymmetry in the light curve of TOI 560b, and the weak velocity evolution during the transit all suggest that stellar winds and/or magnetic fields shape the outflow properties of these planets.  In the planetary frame, the observed velocity evolution represents a redshift early in the transit and a blueshift late in the transit, which is expected if the outflow emerging from the planet's dayside is turned around by the stellar wind and shaped into an increasingly accelerated tail.  However, the redshift seen for TOI 560b suggests that the stellar wind in this system may be weaker than expected or that magnetic fields are significant, as our 3D models including winds all predict a blueshift \citep{zhang_2021c}.  The lack of post-egress absorption for all but one planet argues for similar conclusions.  More 3D simulations accounting for the effects of both the stellar wind and magnetic fields, similar to e.g. \cite{owen_2014,khodachenko_2015,arakcheev_2017}, will be necessary to translate these qualitative statements into quantitative constraints on the stellar wind properties, and to explore how much the stellar wind might contributed to the observed line width.

Despite these caveats, the width of the helium absorption that we observe strongly hints that the outflows are driven by photoevaporation.  However, a better theoretical understanding of the mass loss process and of the properties of these particular exoplanet systems is required before we can make a definitive statement.

\subsection{On masses and eccentricities}
Masses and eccentricites are critical to the interpretation of the helium data.  Under certain conditions, such as a strong background magnetic field compared to the surface magnetic field, the mass loss rate could decline exponentially with planet mass at a rate of $\dot{m} \propto e^{-b}$ where $b = \frac{GM_p}{c_s^2 R_p}$  \citep{adams_2011}.  Under these conditions, a $2.4 M_\Earth$ increase in the planet mass causes a factor of two decrease in the mass loss rate.  \cite{adams_2011} find, however, that when the background field is much weaker than the surface field, the mass loss rate scales as $\dot{m} \propto b^3 e^{-b}$, so that the same change in mass only changes the mass loss rate by several percent.

The eccentricity--or more precisely, the radial velocity of the planet at transit center, $K\sin{\omega}$--is critical to interpreting whether the outflow is blueshifted or redshifted.  Taking TOI 2076b as an example, with an orbital velocity of $\sim$110 km/s, an eccentricity of 0.06 is enough to match the apparent blueshift of $6.7 \pm 1.4$ km/s.  This eccentricity is entirely reasonable; \cite{mills_2019} obtained a mean eccentricity of 0.05 for systems with multiple transiting planets, and 0.21 for systems with a single transiting planet.

Masses, eccentricities, and arguments of periastron can be obtained by detecting the radial velocity signature of the planets with a sufficiently high signal-to-noise ratio.  This is difficult for our targets because of the youth of their host stars, but not impossible.  For example, TOI 560b has a measured mass, and its measured radial velocity during transit enhances rather than diminishes the apparent redshift \citep{zhang_2021b}.  We are collaborating with another team in radial velocity followup of TOI 1430.01, and they have recently obtained a tentative upper limit comparable to our assumed mass of 7 $M_\earth$ (private communication, Joseph Murphy).  Follow-up efforts for TOI 2076 and 1683.01 are underway as part of the TESS Follow-up Program.  For TOI 2076b, which is near a 3:5 period ratio with 2076c and could be in resonance, transit timing variations could also be used to constrain its mass and eccentricity.  Our efforts to model the system's current sample of TTV measurements (Appendix \ref{sec:appendix_ttvs}) were not sufficient to break the mass-eccentricity, but it is possible that future transit observations will, especially if they successfully detect the chopping signal, which our models indicate could have an amplitude of several minutes.

\section{Conclusion}
\label{sec:conclusion}
We surveyed four nearby mini Neptunes orbiting young K-type stars, and detected helium absorption from all of them.  The helium absorption signals show some consistent properties across the four planets, but each planet has unique features as well.  Our simple models suggest that the observed outflows velocities and temperatures appear to be consistent with expectations for photoevaporation, with an inferred mass loss rate within a factor of a few of hydrodynamical predictions. The mass loss timescales for the observed outflows are on the order of several hundred Myr.  This suggests that these mini Neptunes will all likely lose their hydrogen/helium envelopes, evolving into super Earths.

If we wish to develop a more detailed understanding of mass loss processes on these planets, there are several promising paths forward.  On the observational side, additional helium observations of these planets are needed in order to map out the kinematic structures of their outflows in more detail, mitigate the effect of stellar variability on the measured helium signals, and characterize the magnitude of any time variability in the outflow properties.  Ly$\alpha$ observations could be used to probe neutral hydrogen absorption far from the planet and at high velocities, providing complementary constraints on the outflow properties, although many factors could result in the non-detection of even a strong outflow.  If we can measure the planetary masses for TOI 1430.01, 1683.01, and 2076b via radial velocities or transit timing variations, it would also greatly facilitate our theoretical understanding of their outflows, because it would help reduce the dimensionality of the model parameter space and minimize degeneracies with other model parameters, such as the assumed EUV flux and atmospheric composition.  By expanding the sample of young mini Neptunes with outflows detected in the helium line, we can also obtain a better statistical picture of their outflow properties.  This will allow us to search for correlations between specific features (like redshifted/blueshifted absorption) and other planet properties.  We are currently pursuing all of these approaches.  For example, we have HST programs to look for Ly$\alpha$ absorption from TOI 560b and 560c, with plans to write more proposals if these programs return promising data.

On the theoretical front, there is also much work to be done.  In our previous studies of the mini-Neptunes HD 63433b/c \citep{zhang_2021b} and TOI 560b \citep{zhang_2021c}, we modeled our observations with the 1D hydrodynamic code TPCI \citep{salz_2015} and the 3D hydrodynamic code Microthena \citep{wang_2018}, and found that neither code was able to to fully reproduce the observed signals.  We encourage other teams with 3D hydrodynamic codes to try modeling the four outflows reported here.  It is also important to consider whether it is realistic to expect any model to match the observations, given the uncertainties in the composition of the outer atmosphere, the amount of turbulence in the outflow, the XUV flux, the planetary and interplanetary magnetic fields, the velocity of the stellar wind, and the potentially large time variability in all of these quantities due to stellar activity.  If this variability turns out to be a limiting source of uncertainty, expanding the sample of mini Neptunes with helium measurements would help to mitigate against these effects by enabling broader population-level comparisons between data and model as well as an empirical search for correlations between outflow properties and planetary or stellar parameters.

\textit{Software:}  \texttt{numpy \citep{van_der_walt_2011}, scipy \citep{virtanen_2020}, matplotlib \citep{hunter_2007}, dynesty \citep{speagle_2019}, SAS}, exoplanet \citep{exoplanet:joss,exoplanet:zenodo}, PyMC \citep{exoplanet:pymc3}, theano \citep{exoplanet:theano}, celerite2 \citep{exoplanet:foremanmackey18}

All the {\it TESS} data used in this paper can be found in MAST: \dataset[10.17909/b5pm-bb37]{http://dx.doi.org/10.17909/b5pm-bb37}.

\newpage
\acknowledgments
The helium data presented herein were obtained at the W. M. Keck Observatory, which is operated as a scientific partnership among the California Institute of Technology, the University of California and the National Aeronautics and Space Administration. The Observatory was made possible by the generous financial support of the W. M. Keck Foundation.

Based on observations obtained with XMM-Newton (observation ID 0882870701 for TOI 1430, 0882870501 for TOI 1683), an ESA science mission with instruments and contributions directly funded by ESA Member States and NASA.  The grant number is 80NSSC22K0742.

Funding for the TESS mission is provided by NASA's Science Mission Directorate.  We acknowledge the use of public TESS data from pipelines at the TESS Science Office and at the TESS Science Processing Operations Center.  This research has made use of the Exoplanet Follow-up Observation Program website, which is operated by the California Institute of Technology, under contract with the National Aeronautics and Space Administration under the Exoplanet Exploration Program.  This research has made use of the Exoplanet Follow-up Observation Program (ExoFOP; DOI: 10.26134/ExoFOP5) website, which is operated by the California Institute of Technology, under contract with the National Aeronautics and Space Administration under the Exoplanet Exploration Program

\newpage
\appendix

\section{TESS light curves}
\label{sec:tess_light_curves}

\begin{figure*}[h]
  \centering 
  \subfigure {\includegraphics[width=0.5\textwidth]{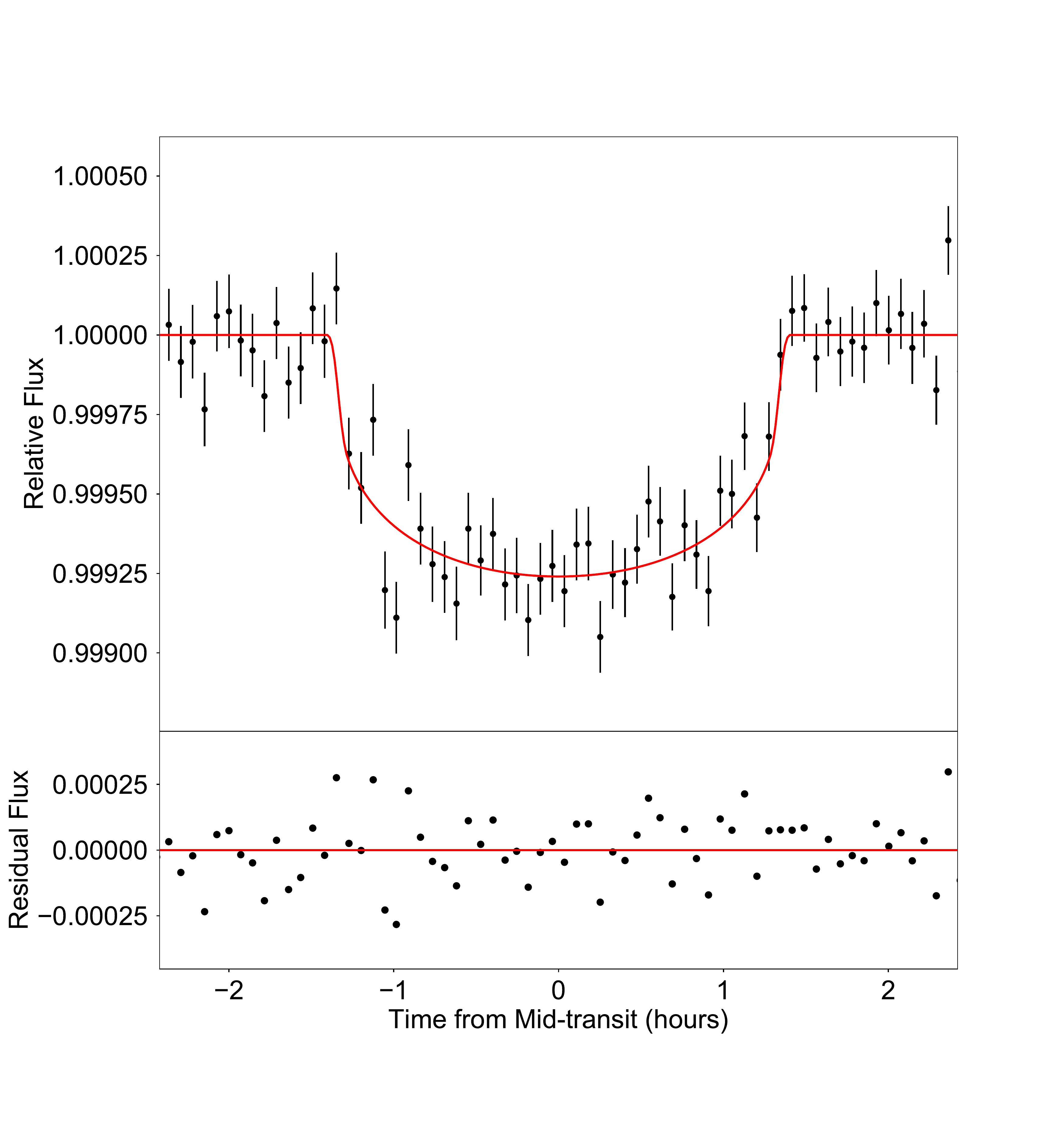}}\subfigure {\includegraphics[width=0.5\textwidth]{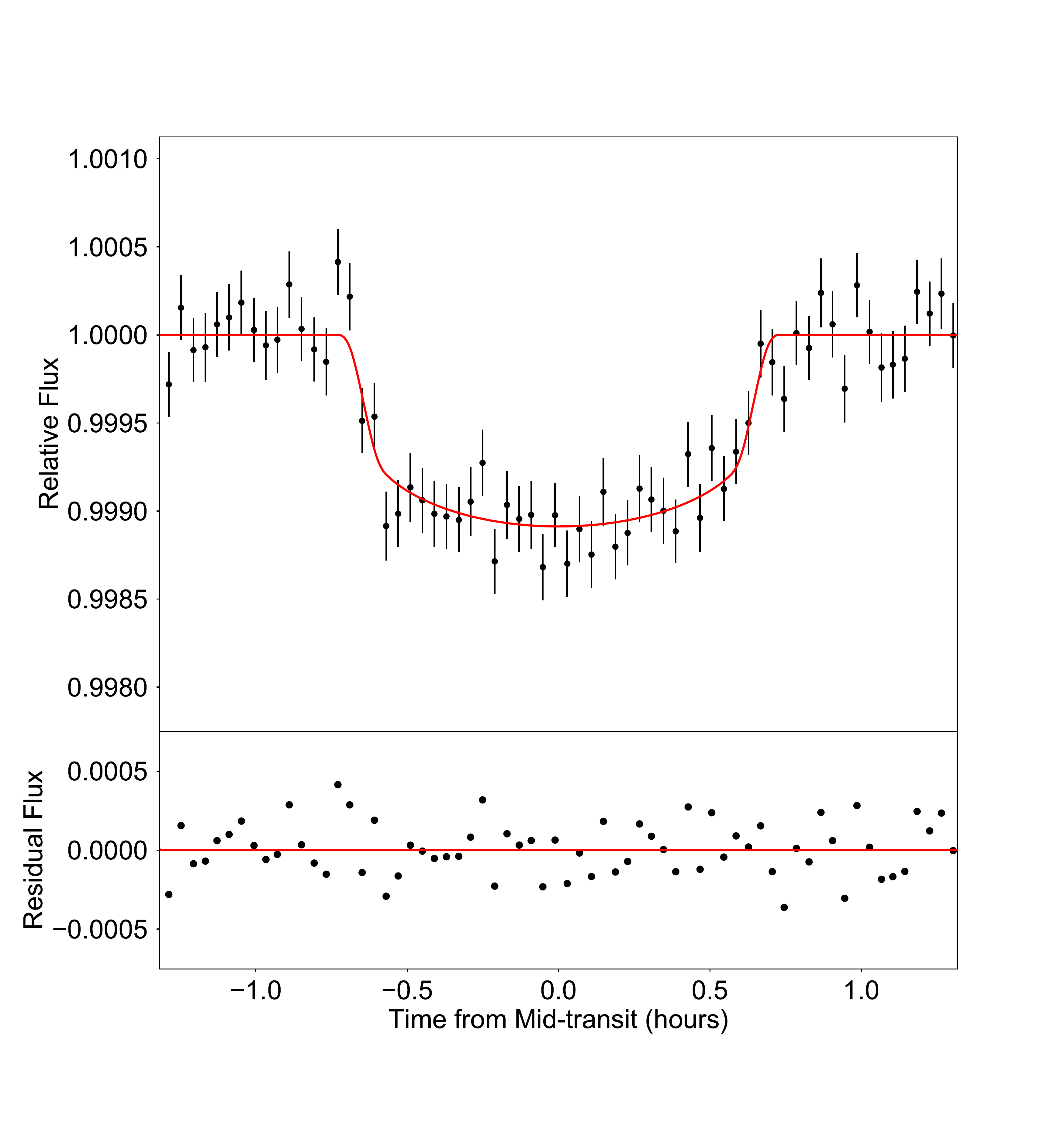}}
  \caption{Phase-folded TESS light curves for TOI 1430.01 (left) and TOI 1683.01 (right).}
  \label{fig:tess_lcs}
\end{figure*}

\section{XMM-Newton X-ray measurements}
\label{sec:appendix_xrays}
\begin{figure*}[h]
  \centering 
  \subfigure {\includegraphics[width=0.5\textwidth]{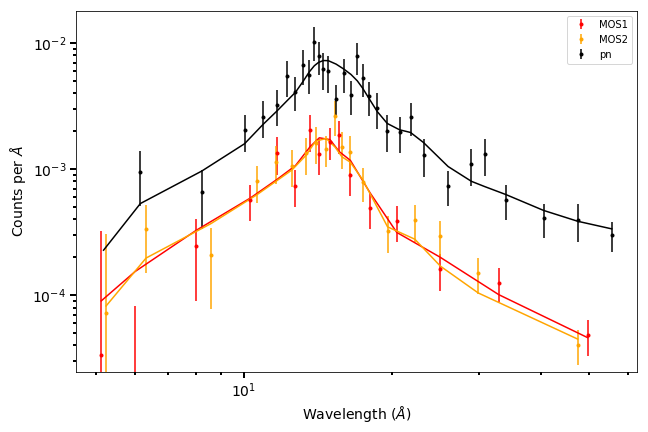}}\subfigure {\includegraphics[width=0.5\textwidth]{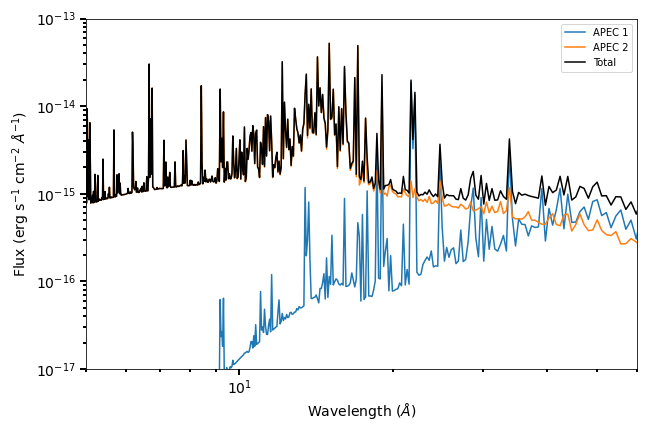}}
  \subfigure {\includegraphics[width=0.5\textwidth]{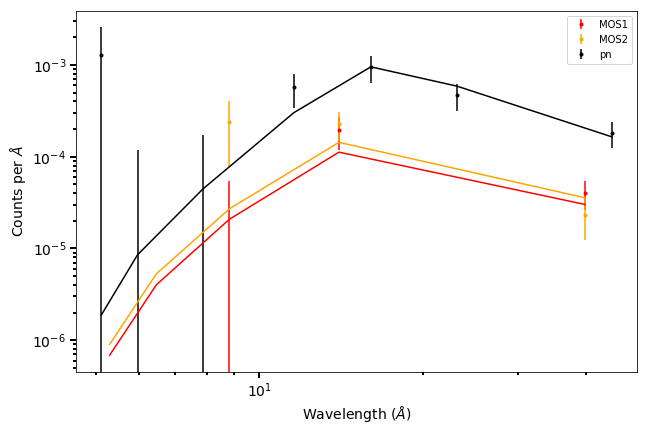}}\subfigure {\includegraphics[width=0.5\textwidth]{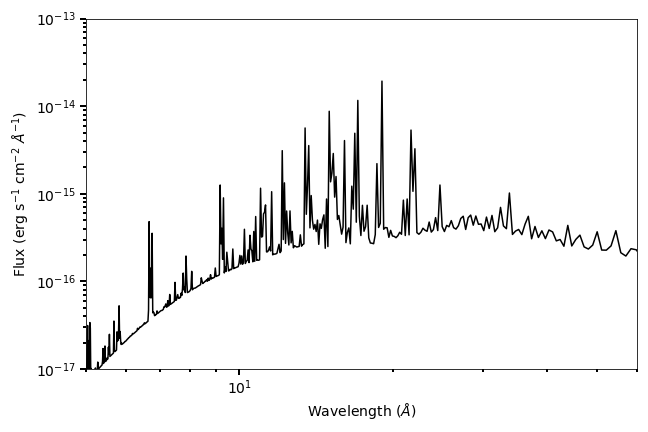}}
  \caption{Left: X-ray spectrum of TOI 1430 (top) and TOI 1683 (bottom), measured by XMM-Newton's three EPIC cameras.  Note that both the throughput and the line spread profile are heavily dependent on wavelength.  Right: the theoretical model that best explains the measurements.  The model consists of two components (top, TOI 1430) or one component (bottom, TOI 1683) of optically thin, collisional plasma in equilibrium.}
  \label{fig:xmm_spectra}
\end{figure*}

\begin{figure*}[h]
  \centering 
  \subfigure {\includegraphics[width=0.5\textwidth]{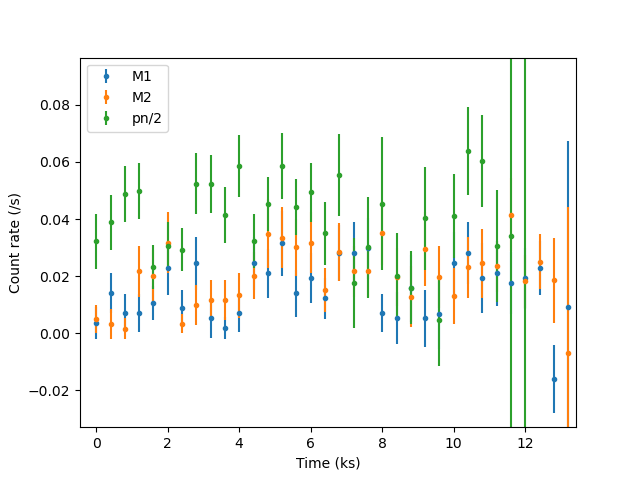}}\subfigure {\includegraphics[width=0.5\textwidth]{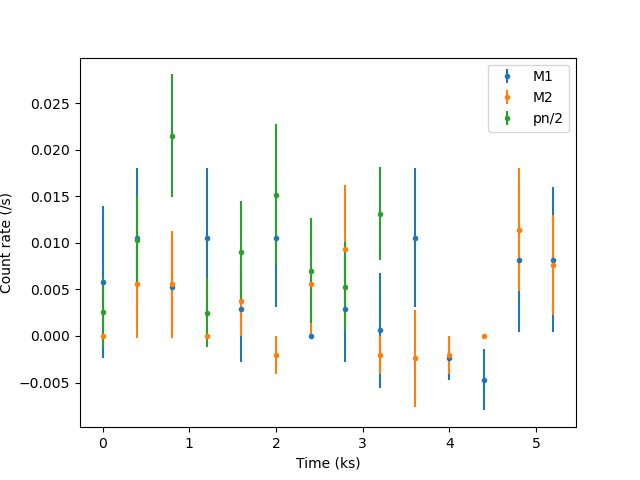}}
  \caption{Background-subtracted X-ray count rate as measured by XMM-Newton's EPIC cameras.  Left: TOI 1430; right: TOI 1683.}
  \label{fig:xmm_lcs}
\end{figure*}

As part of this paper, we analyzed the X-ray measurements of TOI 1430 and 1683 taken by XMM-Newton's three EPIC cameras.  The light curves are plotted in Figure \ref{fig:xmm_lcs}.  The measured spectrum of TOI 1430 and the theoretical model that best explains it are plotted in Figure \ref{fig:xmm_spectra}.  We fit a one-component model for TOI 1683 and a two-component model for TOI 1430, because the SNR is far higher for the latter than for the former.

\begin{table}[ht]
  \centering
  \caption{Model parameters for XMM-Newton data}
  \begin{tabular}{c C C}
  \hline
  	  Parameter & \text{Value (TOI 1430)} & \text{Value (TOI 1683)}\\
      \hline
      Metallicity & 0.29_{-0.16}^{+0.03} & 0.13_{-0.07}^{+0.35}\\
      kT$_1$ (keV) & 0.16_{-0.02}^{+0.31} & 0.27_{-0.03}^{+0.05}\\
      EM$_1$ (cm$^{-3}$) & 8.1_{-2.1}^{+6.6} \times 10^{50} & 17_{-11}^{+7} \times 10^{50}\\
      kT$_2$ (keV) & 0.68_{-0.04}^{+0.06} &\\
      EM$_2$ (cm$^{-3}$) & 14_{-4}^{+8} \times 10^{50} & \\
      Flux$^*$ (erg/s/cm$^2$) & 1.37_{-0.06}^{+0.15} \times 10^{-13} & 0.32_{-0.04}^{+0.05} \times 10^{-13}\\
      \hline
  \end{tabular}
  \tablecomments{$^*$Derived, not a fit parameter.  For the range 5--100 \AA{} (0.124--2.48 keV).}
  \label{table:xmm_params}
\end{table}

The physical parameters we obtained from our MCMC fits to the X-ray data are shown in Table \ref{table:xmm_params}.

\section{Transit timing variations in TOI 2076b}
\label{sec:appendix_ttvs}

We proposed for our TOI 2076b helium observations using the \cite{hedges_2021} ephemeris, but \cite{osborn_2022} subsequently discovered transit timing variations in the three-planet system.  As a result, the \cite{hedges_2021} ephemeris is far less accurate than its error bars imply.  Since \cite{osborn_2022}, TESS has re-observed the system and captured three more transits of b, one more transit of c, and one more transit of d.  The three transits of b are the three immediately following our Keck/NIRSPEC transit.

Accurately predicting the transit time for our Keck observation is critical to the interpretation of the helium outflow.  Ideally, we would use a Bayesian method to model the TTVs of all three planets simultaneously, and use the ensemble of TTV models consistent with the data to generate predictions of the transit time.  However, \cite{osborn_2022} reports that none of their three models are robust: one relies on the dubious assumption that the planets are not in resonance, and the other two arrive at either implausibly low or implausibly high values for the planetary masses, depending on our choice of priors.  To estimate the transit time, we therefore consider three possible approaches: first, we fit a linear ephemeris to the three transits in the recent TESS sector and extrapolate back one epoch; second, we fit a linear ephemeris plus a sinuisoidal TTV model to all transit times for b; third, we fit a full TTV model using the ensemble of measured transit times for all three planets.

All three approaches require us to derive transit times from TESS data.  We do this using the \texttt{exoplanet} package \citep{exoplanet:joss, exoplanet:zenodo} by adapting their tutorial entitled ``Simultaneous Fitting of a Transit with Stellar Variability''.  Specifically, we download the Simple Aperture Photometry (SAP) fluxes and model the stellar variability using the quasiperiodic RotationTerm kernel from the \texttt{celerite2} package \citep{exoplanet:foremanmackey18}, which is a sum of two simple harmonic oscillator terms: a primary term at the stellar rotation period, and a less coherent (lower quality factor, Q) secondary term at half the stellar rotation period.  Unlike in the tutorial, we set the priors on log(Q) and log($\Delta$Q) to be Gaussians with a mean of 12 and standard deviation of 5, instead of a mean of 0 and standard deviation of 2, because visual examination of the light curve shows highly coherent rotation across the entire month for all sectors of observation.  

In addition, unlike in the tutorial, we fix all stellar parameters and all transit parameters except the transit times to the median values derived by \cite{osborn_2022}.  This is for two reasons: first, \cite{osborn_2022} includes non-TESS photometry for all planets, so their values may be more robust; second, we wanted to minimize the number of free parameters to avoid getting caught in a local minimum, which sometimes happened when we experimented with additional parameters.  Our final model includes seven free parameters describing the Gaussian Processes model (mean, white noise level, rotation period, rotational variability $\sigma_{\rm rot}$, log(Q), log($\Delta$Q), relative amplitude of secondary term), in addition to individual transit times for each planets (14 transits in total, 5 in the new sector).  We use PyMC \citep{exoplanet:pymc3} to perform No-U-Turn Hamiltonian Monte Carlo sampling with 2 chains with 1500 tune-up steps and 1000 sampling steps each, and use the built-in tools to check for convergence.  Below are the transit times we measure from our Gaussian Processes fit; non-TESS times from \cite{osborn_2022} are added for completeness:

\begin{table*}[ht]
  \centering
  \caption{Transit times for TOI 2076 planets}
  \begin{tabular}{c c C}
  \hline
  	  Planet & Epoch & {\rm BJD}_{\rm TDB} - 2,457,000 \\
      \hline
      b & 0  & 1743.7169 \pm 0.0035\\
      b & 1  & 1754.0767 \pm 0.0020\\
      b & 18 & 1930.1231 \pm 0.0015\\
      b & 19 & 1940.4811 \pm 0.0025\\
      b & 20 & 1950.8339 \pm 0.0016\\
      b & 57 & 2333.9547 \pm 0.0024\\
      b & 89 & 2665.3392 \pm 0.0024\\
      b & 90 & 2675.6931 \pm 0.0023\\
      b & 91 & 2686.0496 \pm 0.0017\\
      c & 0  & 1748.6931 \pm 0.0010\\
      c & 9  & 1937.8229 \pm 0.0011\\
      c & 25 & 2274.08398 \pm 0.00079\\
      c & 44 & 2673.3653 \pm 0.0011\\
      d & 0  & 1762.6666 \pm 0.0020\\
      d & 5  & 1938.2920 \pm 0.0021\\
      d & 17 & 2359.789 \pm 0.022\\
      d & 18 & 2394.9236 \pm 0.0015\\
      d & 26 & 2675.9326 \pm 0.0016\\
      \hline
  \end{tabular}
  \label{table:transit_times}
\end{table*}

When we use only the new sector of TESS data, compute a linear ephemeris from its 3 transits of b, and extrapolate one epoch back to the Keck observation epoch, we obtain a transit time of $2654.9837 \pm 0.0015$ (BJD\textsubscript{TDB} - 2,457,000).  When we compute all transit times of all planets from all TESS data, add in the non-TESS timings reported by \cite{osborn_2022}, and perform a linear least-squares fit of the b transit times to a linear ephemeris plus a sinuisoid, we obtain a transit time of 2654.9841.  The sinuisoid is assumed to have a superperiod of $1/|2/P_c - 1/P_b| = 713.74 \pm 0.14$ d, consistent with the $713.1 \pm 2.7$ d superperiod found by \cite{osborn_2022}.  For our third method, we combine the N-body code TTVFast \citep{deck_2014} with the nested sampling code \texttt{dynesty} \citep{speagle_2019} to run two full TTV models on all transit times of all planets, following the methodology of \cite{hadden_2016}.  Briefly, we assume zero mutual inclination and have five parameters per planet (mass, ecos($\omega$), esin($\omega$), period, mean anomaly).  The mean anomaly is reparameterized as the epoch 0 transit time, to minimize the degeneracy with period.  In the ``low-mass'' TTV model, we adopt a log-uniform prior on the masses from 1--20 $M_\Earth$ and a uniform prior on the eccentricities from 0.001--0.2; in the ``high-mass'' TTV model, we adopt a uniform prior on the masses and a log-uniform prior on the eccentricities, with the same bounds.  After nested sampling finishes, we draw randomly from the samples, taking into account the weights, and rerun TTVFast to predict the transit time for b at the desired epoch.  We obtain $2654.9830 \pm 0.0024$ for the low-mass model, and $2654.9837 \pm 0.0019$ for the high-mass model.  Since all of our estimates for the transit time are within 1.6 minutes of each other, we are confident that we correctly calculated the transit time despite the TTVs.  We adopt as our final answer the result of the first method, $2654.9837 \pm 0.0015$.  Unfortunately, this timing is 51 minutes earlier than what we expected when we proposed for the Keck observations.  In combination with the telescope and weather problems, this meant that we were unable to observe the first 2/3 of transit.

\begin{figure*}[h]
  \centering 
  \subfigure {\includegraphics[width=0.5\textwidth]{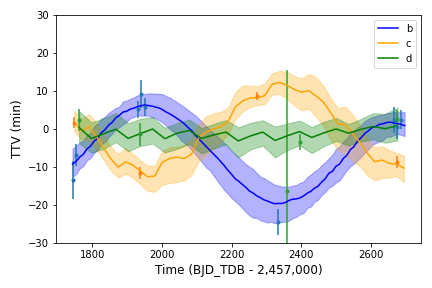}}\subfigure {\includegraphics[width=0.5\textwidth]{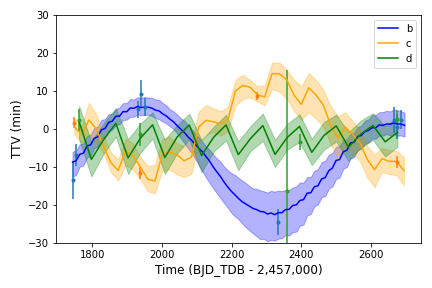}}
  \caption{TTVs computed with the above ephemeris.  The shaded regions are the model predictions from the low-mass nested sampling run (left), or the high-mass nested sampling run (right).}
\label{fig:ttvs}
\end{figure*}

In running the TTV models, we had hoped to compute not just the transit time corresponding to the Keck observations, but the masses of the planets.  Unfortunately, our fits were underconstrained and we found that our masses were highly sensitive to our choice of priors.  The best fit TTV models using both priors are shown in Figure \ref{fig:ttvs}.  We obtained $1.8_{-0.6}^{+1.9}$, $3.1_{-1.0}^{+2.5}$, and $2.3_{-1.0}^{+2.8} M_\Earth$ using the low-mass prior, and $11.3_{-4.2}^+{5.1}$, $11.4_{-2.7}^{+2.5}$, and $6.4_{-4.3}^{+6.6} M_\Earth$ using the high-mass prior.  The low-mass fit favors eccentricities of a few percent, while the high-mass fit favors eccentricities of a few tenths of percent.  The large uncertainties on the masses, together with the strong dependence on the choice of prior, mean that additional transit and radial velocity observations are needed to fully characterize this system.  

As a final step, we improved upon the ephemerides of \cite{osborn_2022} by deriving new periods and transit midpoints.  For systems with TTVs, there is no single correct definition of period.  Possible definitions include: (1) the coefficient found from a linear least-squares fit of transit time vs. epoch; (2) the coefficient found from a fit of transit time vs. epoch of the form $T=PE + T_0 + A\sin{(\omega_{\rm super}PE}) + B\cos{(\omega_{\rm super}PE)}$; (3) the period implied by the position and velocity of the planet at a given time if the other planets didn't exist.  Definitions 1 and 2 are the same if the TTVs are perfectly sinuisoidal and if the observations evenly sample the superperiod.  Definition 3 depends on the time chosen.  For our TTV fits, we choose a time of 1743, just before the first transit of b.  Below, we present the periods and mean transit times that follow from definition 1:

\begin{table*}[ht]
  \centering
  \caption{Linearized ephemerides for TOI 2076b planets}
  \begin{tabular}{c C C}
  \hline
  	  Planet & P (d) & T_0 ({\rm BJD}_{\rm TDB} - 2,457,000) \\
      \hline
      b & 10.355183	\pm 0.000065 & 2178.6439 \pm 0.0023\\
      c & 21.01544 \pm 0.00027 & 2147.9853 \pm 0.0041\\
      d & 35.12561 \pm 0.00011 & 2289.5491 \pm 0.0011\\
      \hline
  \end{tabular}
  \label{table:ephemerides}
\end{table*}

\newpage
\bibliographystyle{apj} \bibliography{main}

\end{document}